\documentclass[a4paper,12pt]{article}
 
\usepackage{bbm}
\usepackage[utf8]{inputenc}
\usepackage{placeins}
\usepackage{amssymb,amsmath,amsfonts}
\usepackage[colorlinks,linkcolor=purple,citecolor=teal]{hyperref}
\usepackage{dsfont}
\usepackage{color}
\usepackage{graphicx}
\usepackage{hyphenat}
\usepackage{wrapfig}
\usepackage{empheq}
\usepackage{textcomp}
\usepackage[caption=false]{subfig}
\usepackage{rotating}
\usepackage{wrapfig}
\usepackage{pdfpages}
\usepackage{verbatim}
\usepackage{setspace}
\usepackage{array}
\usepackage{caption}
\usepackage[pdf]{pstricks}
\usepackage{upgreek}
\usepackage{cmll}
\usepackage{latexsym}
\usepackage{braket}
\usepackage{cite}
\usepackage{epsfig}
\usepackage[left=2.4cm,top=3.3cm,right=2.4cm,bottom=3.3cm,bindingoffset=0cm]{geometry}
\usepackage[titletoc,page]{appendix}
\usepackage{multicol}
\usepackage{youngtab}

\usepackage[normalem]{ulem}     

\newcommand{\beq}{\begin{eqnarray}}
\newcommand{\eeq}{\end{eqnarray}}
\newcommand{\bea}{\begin{eqnarray}}
\newcommand{\bqa}{\begin{eqnarray}}
\newcommand{\eea}{\end{eqnarray}}
\newcommand{\be}{\begin{equation}}
\newcommand{\ee}{\end{equation}}
\newcommand{\diff}{\mathrm{d}}
\newcommand{\tr}{\mathrm{tr}}

\newcommand{\rmc}{\mathrm{c}}

\def\brc{\langle}
\def\ckt{\rangle}

\def\nn{\nonumber}

\def\Tr{\qopname\relax o{Tr}}

\numberwithin{equation}{section}

\numberwithin{equation}{section}

\begin{document}

\title{Anomalies and Dynamics in Strongly-Coupled Gauge Theories, New Criteria for Different Phases, and  a Lesson from Supersymmetric Gauge Theories  }  

\vskip 40pt  
\author{  
Kenichi Konishi$^{(1,2)}$$^*$, Stefano Bolognesi$^{(1,2)}$,  Andrea Luzio$^{(3,2)}$    \\[13pt]
{\em \footnotesize
$^{(1)}$Department of Physics ``E. Fermi", University of Pisa,}\\[-5pt]
{\em \footnotesize
Largo Pontecorvo, 3, Ed. C, 56127 Pisa, Italy}\\[2pt]
{\em \footnotesize
$^{(2)}$INFN, Sezione di Pisa,    
Largo Pontecorvo, 3, Ed. C, 56127 Pisa, Italy}\\[2pt]
{\em \footnotesize
$^{(3)}$Scuola Normale Superiore,   
Piazza dei Cavalieri, 7,  56127  Pisa, Italy}\\[2pt]
\\[1pt] 
{ \footnotesize    kenichi.konishi@unipi.it,    \,   stefano.bolognesi@unipi.it,  \,  andrea.luzio@sns.it}  
} 

\date{}

\vskip 6pt

\maketitle

\begin{abstract}

We review  recent developments in our understanding of the dynamics of  strongly-coupled chiral  $SU(N)$ 
gauge theories in four dimensions, problems which are potentially important in our quest to go beyond the standard $SU(3)_{QCD} \times (SU(2) \times U(1))_{GWS}$   model  
of the fundamental interactions. The generalized symmetries and associated new 't Hooft anomaly-matching constraints allow us to exclude, in a wide class of chiral gauge theories,  
confining vacuum with full flavor symmetries supported by a set of color-singlet massless composite fermions. 
The color-flavor-locked dynamical Higgs phase, dynamical Abelianization 
or more general symmetry breaking phase,  appear as plausible IR dynamics,  depending on  the massless matter fermions present.  We revisit and discuss critically 
several well-known confinement criteria in the literature, for both chiral and vectorlike gauge theories, and propose tentative, new criteria for discriminating different phases.  Finally, we review an idea
which might sound rather surprising at first, but is indeed realized in some softly-broken supersymmetric theories,  that    confinement in QCD is a small deformation 
(in the IR end of the renormalization-group flow) of   a  strongly-coupled, nonlocal,  nonAbelian   conformal fixed point.

\end{abstract}

\vskip 12pt

\hrule

\vskip 6pt

{\footnotesize  * Speaker.    Invited talk at  the Corfu Summer Institute 2023 "School and Workshops on Elementary Particle Physics  and Gravity" (CORFU2023),  
  Corfu, Greece  }

\newpage

\tableofcontents

\bigskip

\newpage

\section{Introduction}
  
The main topic of this talk is the dynamics of strongly-coupled chiral gauge theories in four dimensions.  
This talk is based on several  recent works \cite{BKS}-\cite{BKLproc} and on some earlier ones \cite{AGK}-\cite{GiacomelliK3}.  

 There are several good motivations for these efforts: 
\begin{description}
  \item[(1)]   We are living in a world with a nontrivial chiral structure (e.g., chiral DNA spirals);  
  \item[(2)]   At a microscopic level, the standard model of the fundamental interactions
     $SU(3)_{QCD} \times SU(2)_L \times U(1)_Y$ gauge theory is a chiral theory; 
  \item[(3)]  Possible GUT extensions of the standard model are all chiral. 
\end{description}
Although  the $SU(2)_L \times U(1)_Y$    Glashow-Weinberg-Salam model  is  a chiral gauge theory,   it is weakly coupled and  is well understood in  perturbation theory.  But  it also means that 
it is at best a very  good low-energy effective theory.

 Surprisingly little is known today about the phases of  strongly-coupled chiral gauge theories, after many years of studies \cite{Raby:1979my}-\cite{THooft}. 
This is to be contrasted with the case of vectorlike gauge theories, where we have a far better grasp of the strong gauge dynamics. Quantum Chromodynamics  (QCD)  has been under intense theoretical and experimental investigations over $50$ years, with many solid results established.   Also many exact results are known about the gauge dynamics of ${\cal N}=2$ supersymmetric gauge theories (which are all vectorlike), since the discovery of the 
Seiberg-Witten solutions \cite{SW1,SW2} \cite{curves1}-\cite{Tachikawa}. 

We propose thus the problem:  

~~~

\noindent  {\it Understand better the dynamics of strongly-coupled (especially,  chiral) gauge theories}
   
 ~~~

 \noindent  as a challenge to all theoretical physicists.  
We take as our theoretical laboratories  - or the battle ground -  the following classes of $SU(N)$  gauge theories,    with fermions in various  (anomaly-free) representations:

\begin{description}
  \item[(i)]    $  \psi^{ij} \;,  \eta_i^B\;,\quad    (i,j, = 1,2,\ldots,N\;,  B=1,2,\ldots, N+4), $    that is, 
    \be      
\yng(2) \oplus   (N+4) \, {\bar{{\yng(1)}}} \,      \label{BYmodels} 
\ee
plus possible $p$ pairs of Dirac fermions in the fundamental representations  
(known as the generalized  Bars-Yankielowicz  (BY)    models), 

  \item[(ii)]  $\chi_{[ij]}\;,  {\tilde \eta}^{Bj}\;,\quad   B=1,2,\ldots, (N-4)$,  
    \be       \yng(1,1) \oplus   (N-4) \,{\bar   {{\yng(1)}}}\,           \label{GGmodels}    
      \ee
 plus possible $p$ pairs of Dirac fermions in the fundamental representations   (called often as the generalized  Georgi-Glashow (GG)  models),

  \item[(iii)]     $\psi^{\{ij\}} \;,    \chi_{[ij]}\;,     \eta_i^A\;,  \quad   A=1,2,\ldots, 8\;,  $
  \be     \yng(2) \oplus  {\bar  {\yng(1,1)}}  \oplus   8  \   {\bar  {\yng(1)}}\;.  
\ee
(sometimes referred to as the $\psi\chi\eta$ model), 

  \item[(iv)]     $   \tfrac{N- 4}{k}   \,  \psi^{\{ij\}}  \oplus   \tfrac{N+4}{k} \,  {\bar  \chi_{[ij]} }$
    \be   \frac{N-4}{k}  \,    \yng(2)   \oplus   \frac{N+4}{k}     \,    {\bar  {\yng(1,1)} }  \ ,   \label{tensors}
\ee

    \item[(v)]   $\psi$'s in a self-conjugate, antisymmetric representation, (e.g., for $SU(6)$),  
         \be       \yng(1,1,1)     \;,  \ee
      \item[(vi)]     $N_f$      \, $\eta \oplus {\bar \eta}$'s   \,\, (QCD)
      \item[(vii)]     $N_f$   \,\,  $\lambda$     \,\, (adjoint QCD),       
    
      \end{description}
       and some others, to start with.    

A well-known tool of the analysis - the 't Hooft anomaly matching conditions -, unfortunately is not  sufficiently stringent  \cite{BK},\cite{Appelquist:2000qg}-\cite{Dimopoulos:1980hn}.  
For this reason we  appeal to a  more powerful approach, mainly based on the generalized symmetries and  associated  anomalies \cite{BKL1}-\cite{BKLZ2}. 

\section{Anomalies and dynamics:  phases of chiral gauge theories \label{newmore} }

The main tool of the  following analyses   is the idea of the generalized symmetries  \cite{Seiberg}-\cite{GKKS}.  The first step is to go from the  conventional $0$-form symmetries, acting on local fields, to $k$-form symmetries,
acting on line, surface, etc. operators.  The idea of $1$-form symmetry is actually familiar from the example of the so-called center symmetry in $SU(N)$  Yang-Mills (YM) theory.  It acts on the Polyakov loop, 
\be   e^{i  \oint _{\gamma}  A   }  \to      \Omega_N      e^{i  \oint _{\gamma}  A   } \;, \qquad      \Omega_N =   e^{2\pi i /N}  {\mathbf 1}   \in  {\mathbbm Z}_N\;. 
\ee
The vanishing  (or nonvanishing) of the vacuum expectation value (VEV) 
\be  \brc  e^{i  \oint _{\gamma}  A}  \ckt
\ee
can be used as a criterion of the confinement (or Higgs) phase,   respectively.     Below we shall concentrate on the use of $1$-form  symmetries.

The second step is to  consider the ``gauging" of the  $1$-form discrete ${\mathbbm Z}_N$  symmetry.  
  The gauging of 1-form  discrete  $\mathbb{Z}_{N}$   symmetry
proceeds  by introducing the 2-form gauge fields $(B_\rmc^{(2)}$, $B_\rmc^{(1)}\big)$,   
\be  N  B_\rmc^{(2)} = d  B_\rmc^{(1)}\;,    \label{1fgconstraint}
\ee
and coupling them to   the  $SU(N)$  gauge fields  $a$ 
  appropriately.  
This is done by embedding it  into a  $U(N)$ gauge field $\widetilde{a}$ as 
\be
\widetilde{a}=a+\frac{1}{N}B^{(1)}_\rmc
\ee
and  requiring the  invariance under the 1-form gauge transformation, 
\begin{align}  B_\rmc^{(2)} & \to B_\rmc^{(2)}+\diff \lambda_\rmc\;, \qquad
 B_\rmc^{(1)}  \to B_\rmc^{(1)}+ N  \lambda_\rmc    \;,   \nonumber \\ 
 \widetilde{a} &\to \widetilde{a}+\lambda_\rmc\;,
   \label{werequire} 
\end{align}
where
$\lambda_\rmc$   is the  (1-form) gauge function such that  
 \be       \oint \lambda_\rmc  =  \frac{2\pi \ell}{N}\;,   \qquad \quad  (n  \in {\mathbbm Z}, \quad \ell  \in {\mathbbm Z})\;.
 \ee

The third important step is  the idea of {\it  color-flavor locked}    $1$-form discrete ${\mathbbm Z}_N$  symmetry.  
Consider an $SU(N)$ gauge theory with a set of the massless matter Weyl fermions $\{\psi^k\}$.  In general, 
 the  center  $\mathbbm Z_N $ symmetry is broken by the presence of the  fermions (unless the fermions are all  in the adjoint representation of $SU(N)$). 
 However  the situation changes if  some global, nonanomalous $U(1)$ symmetries, $U_i(1), i=1,2, \ldots$, are there, which
 are gauged (in the usual sense by the introduction of external gauge fields  $A_i^{\mu}$).  In those cases  the color ${\mathbbm Z}_N  \subset SU(N)$ and the $U_i(1)$ transformations can compensate each other, restoring the symmetry.

As one encircles a closed loop  $L$    in spacetime, the fields transform as  
\be  {\cal P}  e^{i \oint_L    a }    \to      e^{ \tfrac{2\pi  i}{N}  }    {\cal P}  e^{i \oint_L  a }  \;; \quad 
          \psi^k  \to   e^{  \tfrac{ 2\pi i  {\cal N}_k  }{N}}    \psi^k   \;, \qquad     {\mathbbm Z}_N  \subset SU(N)\;;  \label{center1}  \ee 
   \be    \Pi_i \, e^{i \oint_L  A_i }  \to   \left(  e^{  2\pi i   \sum_{i,k}   q_k^{(i)} }    \right)     \Pi_i \, e^{i \oint_L  A_i } \;;       \qquad     \psi^k   \to    e^{  2\pi i   \sum_{i,k}  q_k^{(i)} }  \psi^k\;, \qquad U_i(1)\;;  \label{center2} 
   \ee
 where  $a \equiv     a_{\mu}^A  t^A   \; dx^{\mu}$  is the $SU(N)$ gauge field;      ${\cal N}_k$ is the $N$-ality of the $k$th fermion,   $ q_k^{(i)} $ is the charge of  $\psi_k$ under  $U_i(1)$.    The factor  $e^{i \oint_L  A_i } $   is nothing but the Aharonov-Bohm phase for the $i$-th fermion.

   We recall that  the center symmetry is formally defined as  a path-ordered sequence of local  $SU(N)$ gauge transformations along the loop:  the fermions must also  be transformed in order to keep the action invariant.   After encircling the loop and coming back to the original point, the gauge field is transformed by a nontrivial periodicity with a $\mathbbm Z_N $  factor,  dragging  the  fermions fields to transform as in (\ref{center1}).  It would invalidate  their periodic boundary condition  (i.e., their uniqueness at  each spacetime point).
   This is the reason why the presence of a fermion, such as the one in the fundamental representation,   breaks the center symmetry itself  \footnote{In the case of the Polyakov loop defined in the Euclidean spacetime, the fermions are required to satisfy antiperiodic boundary condition,  but the conclusion is the same.}.

  When the conditions 
 \be       \sum_i  q_k^{(i)} = -     \tfrac{{\cal N}_k}{N} \;,    \qquad     \forall  k   
 \ee
are satisfied,  however,    a new, color-flavor locked center symmetry  ({\ref{center1}),   ({\ref{center2}),  can be defined, 
accompanying the color  $\mathbbm Z_N $   center transformations with appropriate  $U_i(1)$ gauge transformations.  

As the ordinary   $\mathbbm Z_N $ center transformation,  such a color-flavor combined $\mathbbm Z_N $ center symmetry  is still just a 
{\it global   1-form symmetry.}

Finally, we put  all these ideas together, and introduce  the {\it  gauging
of the color-flavor locked 1-form symmetry}  and studying possible topological obstructions in doing so (the generalized  't Hooft's anomalies) \cite{Seiberg}-\cite{AnberHong}, \cite{BKL1}-\cite{BKLZ2}.  As in the case of conventional gauging of 0-form symmetries, the idea of gauging is that of {\it  identifying} the field configurations connected by the given symmetry transformations, and of eliminating the double counting.  
However, as one is now dealing with a 1-form symmetry,  the associated gauge transformations are parametrized by a 1-form Abelian  \footnote{Here we remember the crucial aspect of higher form symmetries: they are all Abelian.  This is the reason why the color-flavor locked 1-form symmetries are possible.}    gauge function,
  $\lambda =  \lambda_{\mu}(x)  dx^{\mu}$;  see (\ref{werequire}).

\subsection{$({\mathbbm Z}_2)_F$ anomaly    \label{Z2anomaly} }  

One of the most significant results found   is a  $({\mathbbm Z}_2)_F$ anomaly \cite{BKL1}-\cite{BKLReview},\cite{BKLZ2}:  a quantum anomaly associated with  the nontrivial  classical fermion parity 
symmetry, 
\be    \psi_i \to - \psi_i  \;,   \qquad \forall i\;,    \label{Z2symmetry}  
\ee
found in all BY and GG  models  (see (\ref{BYmodels}),  (\ref{GGmodels})), with even $N$ and even   $p$.   We call these type I  BY and/or GG models, and refer to all others
 (either $N$ or $p$ or both,  odd)  as   type II models.  In the standard quantization,  (\ref{Z2symmetry})  is conserved both classically, and quantum-mechanically, i.e., it is non anomalous. 
  The standard calculation of the possible anomaly 
 (say, \`a la  Fujikawa),  yields
 \be \Delta S =  \sum_i    c_i \times \frac{1}{8 \pi^2} \int_{\Sigma_4} d^4 x  \,\,  \tr  \,   F_{\mu \nu} {\tilde F}^{\mu \nu}  \times (\pm  \pi) = 2\pi {\mathbbm Z}\;, \qquad  
 \ee
where
\be       \sum_i    c_i = 2 {\mathbbm Z} \ne 0\;,  
\ee
and the partition function is invariant.
The fact that   (\ref{Z2symmetry})  is respected,  in the type I models, by the instantons because  the sum of the contributions from different fermions $  \sum_i    c_i$ is a  {\it   nonvanishing, but even,  integer,   and not because it is zero (as in the type II models)},        is fundamental.

In fact,  in type I models, the symmetry group  has a disconnected structure  (before dividing out by a common  ${\mathbbm Z}_N$ factor). For instance,  in the $\psi\eta$ models, it is 
 \be     \frac{  SU(N)_c \times   SU(N+4)   \times  U(1)_{\psi\eta}   \times  {\mathbbm Z}_2  }{{\mathbbm Z}_N}    \;.  \label{group}
    \ee
    The calculation of the ${\mathbbm Z}_N$  anomaly, by   gauging the  color-flavor locked 1-form ${\mathbbm Z}_N$  symmetry  gives, in all type I models,   the result (a master formula)  
    \be    \Delta S^{\rm Mixed ~ anomaly}   =  (\pm \pi)  \cdot  \sum_{\rm fermions} \left(   (d(R)  {\cal N}(R)^2  - N \cdot  D(R) \right)  \, 
    \frac{1}{8 \pi^2}   \int_{\Sigma_4}\,  \left(  B_{\rmc}^{(2)} \right)^2   = \pm \pi\;, \label{master}  
    \ee
    where $d(R)$ is the dimension of the representation $R$,  ${\cal N}(R)$ its $N$-ality, and   $D(R)$ its Dynkin index.   
    The partition function changes sign under   (\ref{Z2symmetry}): it is anomalous  \cite{BKL2}-\cite{BKLReview}.  
    
   In the IR,   the assumed massless  composite fermions   (see  Table~\ref{Confining})
   cannot  support such an anomaly,  as  they are color singlets and not coupled to the  $B_{\rmc}^{(2)}$  field.    
 It means that  the confining, flavor symmetric vacuum contemplated in some earlier literature \cite{Appelquist:1999vs}-\cite{Eichten:1985fs}    
 cannot be dynamically realized in the infrared. 


\begin{table}
 \centering 
  \begin{tabular}{|c|c|c|c|c|}
\hline
 fields  &  $SU(N)_{\mathrm{c}}  $    &  $ SU(N+4)$     &   $ U (1)_{\psi \eta}  $  &  $({\mathbb{Z}}_{2})_F$      \\
 \hline
  \phantom{\huge i}$ \! \!\!\!\!$  $\psi$   &   $ { \yng(2)} $  &    $ (\cdot) $    & $   \frac{N+4}{2}$     &  $+1$      \\
 $ \eta$      &   $  \bar{ {\yng(1)}}  $     &  ${ \yng(1)} $   &   $  - \frac{N+2}{2} $  &  $-1$   \\
 \hline  
 $ \phantom{{\bar{  {\bar  {\yng(1,1)}}}}}\!  \! \!\! \! \!  \!\!\!$  $ {\cal B}^{AB} $      &   $ (\cdot)   $     & $ { \yng(1,1)}$   &    $  - \frac{N}{2} $   &   $-1$     \\
\hline
\end{tabular}
\caption{\footnotesize The charges of UV and IR fermions with respect to the unbroken symmetry groups,    in a putative 
confining, chirally symmetric vacuum, contemplated in some earlier literature.  
  $ {\cal B}^{AB} $ are the hypothetical massless baryons, $\sim \psi \eta^A \eta^B$.  That they satisfy the conventional anomaly-matching conditions  can be checked by
  studying simple arithmetic equations  involving various triangle  diagrams,    $ (SU(N+4))^3$,   $ (SU(N+4))^2-U (1)_{\psi \eta}$,     $(U (1)_{\psi \eta})^3$,  $U (1)_{\psi \eta}-Gravity^2$.      
  See \cite{BKL4} for explicit exposition of these anomaly-matching arithmetics  in all BY and GG models.      } 
\label{Confining}
\end{table}

This conclusion was challenged by Tong and others  \cite{Tong}.    The subtle  point noted already in \cite{BKL2}  is that  the color-flavor locked ${\mathbbm Z}_2$ transformation  corresponds to the background fields winding 
simultaneously    as 
(in the case of the $\psi\eta$  model)   \footnote{$A_0$  is the background gauge field for  the anomaly-free $U_{\psi\eta}(1)$  global symmetry group.}   
\be       \oint  B_\rmc^{(1)} = 2\pi\;,  \qquad   \oint A_0  = \pi\;
\ee
(the latter  corresponding  to  (\ref{Z2symmetry})).  But the latter implies a singular  ${\mathbbm Z}_2$  vortex configuration, a possible technical  (ethical?)   issue.  

The authors of \cite{Tong} propose,  instead, to use a regular  $U_{\psi\eta}(1)$ field $A_0 $ and the consequent color-flavor locked   ($B_\rmc^{(1)},  B_\rmc^{(2)}$) field such that   
\be      \oint A_0  = 2  \pi \;,   \qquad     \oint  B_\rmc^{(1)} = 4\pi\;,        \label{problem}
\ee
i.e.,  with twice  the 't Hooft flux.   Hence  there is  no  ${\mathbbm Z}_2$     anomaly   (see (\ref{1fgconstraint}) and (\ref{master}))!

The problem with this argument is that  the flux  (\ref{problem})  corresponds to a trivial element of  ${\mathbbm Z}_2$  holonomy  group,  
\be    \psi_i \to    \psi_i  \;,   \qquad \forall i\;:     \label{noZ2symmetry}  
\ee
a trivial (i.e., no) transformation. That this  "transformation" is found to be nonanomalous is certainly a good news,  but the significance of such a statement  is not entirely clear \footnote{Another observation made in \cite{Tong} is that   (\ref{Z2symmetry})  is a part of the proper Lorentz group.   Again, this is true and is well known, but is not a point which can be used to try to invalidate the argument of \cite{BKL2}: see Sec.~2.2   bellow.
   }.

The cure for this technical issue (the need to use the singular  ${\mathbbm Z}_2$ vortex like configuration)  of the 
original work on the $({\mathbbm Z}_2)_F$ anomaly \cite{BKL2, BKL4, BKLReview},      can be found \cite{BKLZ2}  by   starting  with a model  with an extra Dirac-like pair of the fermions.  
Consider the $\psi\eta$ model, with an additional  Dirac  pair  $q, {\tilde q}$ fermions,   and a singlet scalar field $\phi$,  with the Yukawa coupling    (let us call it the  ``X-ray model),  
\be     \Delta L = g \,  \phi \,  q\, {\tilde q}    + h.c.\,,  \qquad     \brc \phi \ckt   = v \gg  \Lambda_{\psi\eta}\;.  
\ee
Before the  VEV of  $\phi$ forms  (i.e., in the UV)   the model has a nonanomalousymmetry  
\be        \frac{SU(N)_c \times  SU(N+4) \times  {\tilde U}(1) \times U_0(1) }{{\mathbbm Z}_N}  
\ee
where    the charges   of the fermions  are listed in Table~\ref{SymmetryX}.
	\begin{table}
		\centering
			\begin{tabular}{c|c|c|c|c|c|c}
				 & $SU(N)_c$ & $SU(N+4)$ & $U(1)_{\psi\eta}$ & $U(1)_V$ & $U_0(1)$ & $\tilde U(1)$  \\
				 \hline 
				$\psi$ & $\yng(2)$ & $(\cdot)$ & $\frac{N+4}{2}$ & $0$ & $1$ & $\frac{N+4}{2}$ \\
				$\eta$ & $\bar{\yng(1)}$ & $\yng(1)$ & $-\frac{N+2}{2}$ & $0$ & $-1$ & $-\frac{N+2}{2}$ \\
				\hline
				$q$ & $\yng(1)$ & $(\cdot)$ & $0$ & $1$ & $1$ & $\frac{N+2}{2}$ \\   
				$\tilde q$ & $\bar{\yng(1)}$ & $(\cdot)$ & $0$ & $-1$ & $1$ & $-\frac{N+2}{2}$ \\ 
				$\phi$ & $(\cdot)$ & $(\cdot)$ & $0$ & $0$ & $-2$ & $0$ \\                                
			\end{tabular}
			\caption{\footnotesize The fields and charges of the $X$-ray model with respect to the nonanomalous symmetries.  
			}
			\label{SymmetryX}
		\end{table} 
		Note that this time the color-flavor locked  ${{\mathbbm Z}_N}$  symmetry is  in the intersection among the nonanomalous, {\it  continuous}, symmetry groups,  
		\be    SU(N_c)   \cap    \{   U(1)_{\psi\eta}  \times  U_0(1) \}
		 \ee
		 hence no difficulties arise in introducing the dynamical and background gauge fields associated with these symmetry factors   (cfr.  (\ref{group})).

			The calculation of the mixed anomalies \cite{BKLZ2}   yields 
			\begin{description}
  \item[(i)]     ${\tilde U}(1) -   \left( B_\rmc^{(2)}\right)^2  $ anomaly:
  \be  \delta  S_ { \delta \alpha}  =     \frac {\tilde  C}  {8\pi^2}   \int_{\Sigma_4}   \,  (B_{\rm c}^{(2)})^2  \,  \delta \alpha\;,      \qquad   
    {\tilde  C}    =        -     \frac{N^3 (N+3)}{2}   \ne 0\;.   \label{Xan1} 
  \ee
 The  ${\tilde U}(1)$ symmetry  
   is  broken (i.e., gets anomalous) by the generalized 1-form gauging  of the ${\mathbbm Z}_N$. 
 
  \item[(ii)]    $A_0-(B_{\rm c}^{(2)})^2  $   anomaly:
   
An analogous calculation leads to  the $U_0(1)$  anomaly due to   the  1-form gauging of the ${\mathbbm Z}_N$ symmetry, 
\be  \delta  S_ { \delta \alpha_0}  =      {C_0 \over 8\pi^2}   \int_{\Sigma_4}   \,  (B_{\rm c}^{(2)})^2  \,  \delta \alpha_0\;,   \qquad 
C_0   =    N^2 (N+3) \;.       \label{Xan22} 
 \ee
  This appears to imply that the $U_0(1)$ symmetry is  also  broken by  the  1-form gauging of the ${\mathbbm Z}_N$ symmetry. 
 However,  the scalar VEV   $\brc \phi\ckt = v$  breaks spontaneously  the $U_0(1)$ symmetry  to  ${\mathbbm Z}_2$   (see Table~\ref{SymmetryX}).
It means that, in contrast to  (\ref{Xan1}), the generic  variation $\delta \alpha_0$  
cannot be used in (\ref{Xan22}) to examine the generalized   anomaly-matching  check. For that purpose,  we can use  only
the nonanomalous \footnote{In the sense of the standard strong anomaly.} and unbroken symmetry operation,  i.e.,  variations corresponding  to   
 a nontrivial  ${\mathbbm Z}_2$ transformation  
 $ \delta \alpha_0 =\pm \pi$   (see Table~\ref{SymmetryX}).
   Taking into account the nontrivial 't Hooft flux 
 \be        {1  \over 8\pi^2}   \int_{\Sigma_4}   \,  (B_{\rm c}^{(2)})^2    =        \frac{n }{N^2}\;,  \qquad n \in {\mathbbm Z}\;,       \label{Xan2}   
 \ee
and the crucial coefficient of the anomaly,   
 $C_0=N^2 (N+3)$,   it is seen that  the partition function changes sign,     for even  $N$.  
  We thus reproduce exactly  the ${\mathbbm Z}_2$  anomaly first found  in  \cite{BKL2}.   This anomaly cannot be reproduced (matched)  by the low-energy, massless baryons, as they are not coupled to   $B_{\rmc}^{(2)}$.

\end{description}

To conclude,   the  mixed anomaly   $({\mathbbm Z}_{2})_F  - [{\mathbbm Z}_{N}]^2$  means  that confinement and the full global chiral symmetries (no condensates)  are not  compatible, in type I
BY and GG models: one or both must be abandoned.   The dynamical Higgs phase discussed below   seems to be fully consistent. 

\subsection{Dynamical Higgs phase  \label{sec:Higgs} }  

That the conventional 't Hooft anomaly-matching condition  is  also consistent with a dynamical Higgs phase 
characterized by certain bifermion condensates in the BY and GG models  is well known.     For instance, in the $\psi\eta$ model,  a possible bifermion condensate
is \cite{Appelquist:2000qg}  
\be   \brc  \psi^{\{ij\}}   \eta_j^A  \ckt =  C\,  \delta^{i A} \;,   \quad    i, A =1,2,\ldots, N\;,      \label{DHiggs} 
\ee
which breaks the color and   flavor symmetries  as
\be   G      \to   G^{\prime}  = SU(N)_{CF}  \times  SU(4)_F \times U^{\prime}(1)\;.   
\ee
The low-energy theory is described by a set of massless composite fermions (``baryons") and of massless  (Nambu-Goldstone) bosons.  The baryons listed in Table~\ref{Higgs}
saturate the conventional 't Hooft anomaly-matching conditions with respect to the unbroken symmetry group. 
\begin{table}[h!t]
{
  \centering 
  \begin{tabular}{|c|c|c |c|c|c|  }
\hline
$ \phantom{{{   {  {\yng(1)}}}}}\!  \! \! \! \! \!\!\!$   & fields   &  $SU(N)_{\rm cf}  $    &  $ SU(4)_{\rm f}$     &   $  U(1)^{\prime}   $   &  $({\mathbb{Z}}_{2})_F$    \\
 \hline
   \phantom{\huge i}$ \! \!\!\!\!$  {\rm UV}&  $\psi$   &   $ { \yng(2)} $  &    $  \frac{N(N+1)}{2} \cdot   (\cdot) $    & $   N+4  $   & $1$  \\
 & $ \eta^{A_1}$      &   $  {\bar  {\yng(2)}} \oplus {\bar  {\yng(1,1)}}  $     & $N^2 \, \cdot  \, (\cdot )  $     &   $ - (N+4) $    & $-1$ \\
&  $ \eta^{A_2}$      &   $ 4  \cdot   {\bar  {\yng(1)}}   $     & $N \, \cdot  \, {\yng(1)}  $     &   $ - \frac{N+4}{2}  $   & $-1$  \\
   \hline 
   $ \phantom{{\bar{ \bar  {\bar  {\yng(1,1)}}}}}\!  \! \!\! \! \!  \!\!\!$  {\rm IR}&      $ {\cal B}^{[A_1  B_1]}$      &  $ {\bar  {\yng(1,1)}}   $         &  $  \frac{N(N-1)}{2} \cdot  (\cdot) $        &    $   -(N+4) $     & $-1$ \\
       &   $ {\cal B}^{[A_1 B_2]}$      &  $   4 \cdot {\bar  {\yng(1)}}   $         &  $N \, \cdot  \, {\yng(1)}  $        &    $ - \frac{N+4}{2}$   & $-1$   \\
\hline
\end{tabular}  
  \caption{\footnotesize   The charges of the  UV and IR  fermions with respect to the unbroken symmetry groups, in the color-flavor locked, dynamical Higgs  phase of the $\psi\eta$ model.  
  $A_1$ or $B_1$  stand for  $1,2,\ldots, N$,   $A_2$ or $B_2$ the rest of the flavor indices, $N+1, \ldots, N+4$.  
   }\label{THiggs}
}
\end{table}

Unlike the case of the confining, chiral symmetric vacuum   (Table~\ref{Confining}), in the dynamical Higgs phase here  the conventional  't Hooft anomaly-matching  is
totally obvious, as after the Dirac pair of fermions get massive and decouple, the set of the remaining massless fermions are identical in UV and in IR.  (See Table~\ref{THiggs}.)

And unlike the  case of the confining, chiral symmetric vacuum   (Table~\ref{Confining}), here in the dynamical Higgs phase there is no  difficulty in the matching of the $1$-form,  mixed  't Hooft anomalies,
$({\mathbbm Z}_2)_F  -  ({\mathbbm Z}_N)^2$.    The vacuum breaks spontaneously both  
 $SU(N)_{\mathrm{c}}$ and $ U (1)_{\psi \eta}$.   The  color-flavor locked $1$-form ${\mathbbm Z}_N$   is broken in the IR.

A subtle, possibly confusing point is that  
the 0-form  $({\mathbbm Z}_2)_F$  symmetry itself does not need to be, and indeed is not,  spontaneously broken, as all bifermion condensates are invariant under   
(\ref{Z2symmetry}).   In fact, as the  fermion parity   coincides with an angle $2\pi$ space rotation,  a spontaneous breaking of   $({\mathbbm Z}_2)_F$  would have been a disaster:  the spontaneous breaking of the Lorentz invariance.  Which does not occur.

   In this respect,  even though  the mixed anomaly   $({\mathbbm Z}_{2})_F  - [{\mathbbm Z}_{N}]^2$  found in \cite{BKL2} and in  \cite{BKL4, BKLReview},\cite{BKLZ2},     may  look   at first sight similar  to the mixed anomaly 
   $CP -    [{\mathbbm Z}_{N}]^2$   in the pure $SU(N)$ Yang-Mills theory at $\theta = \pi$  \cite{GKKS},   
   the {\it way} the mixed anomaly manifests
itself at low energies  is different.  In the latter case,   the new anomaly is consistent with,  or implies,   the phenomenon of the double vacuum degeneracy   and the consequent spontaneous $CP$ breaking \`a la Dashen \cite{Dashen} \footnote{That this occurs in $SU(N)$ Yang-Mills theory at   $\theta = \pi$ has been known for some time, from the QCD Effective Lagrangian analysis \cite{DiVecchiaVenez,Witten:1980sp}  and also from soft supersymmetry breaking perturbation \cite{Konishi,Evans} of  the exact Seiberg-Witten solutions \cite{SW1,SW2} of pure ${\cal N}=2$  supersymmetric Yang-Mills theory. Still, it is remarkable  that the same result is reproduced by a symmetry consideration,   based on the generalized mixed-anomaly matching requirement. 
}.
 

\subsection{Strong anomaly and phases}  

In all anomaly-matching  argument \`a la 't Hooft,   conventional or generalized,  we  consider only the symmetries which are anomaly-free, i.e.,  which are not broken by the nonperturbative,  strong-interaction effects.  The breaking of the axial  $U_A(1)$  symmetry in QCD, broken by the instantons,  is a famous example of such a "strong anomaly". 
  Actually,  the strong anomaly should not be considered as a simple loss of a symmetry, but as a particular manifestation of a classical symmetry through the strong dynamics.  Recently 
it was shown   \cite{BKL4} that the consideration of the strong anomaly gives rather a clear indication about the possible phase of  a wide class (BY and GG)  of chiral gauge theories.  

In QCD  (e.g., with $N_F=2$), the global flavor  $SU_L(2) \times SU_R(2) \times U_V(1) \times U_A(1)$    symmetries are  broken by the  biquark condensate,  
\be    \brc U \ckt =   \brc  {\bar {\psi_R}} \psi_L   \ckt \sim \Lambda  \ne 0\;   \label{VEV}
\ee
to   $SU_V(2) \times U_V(1)$.  For small quark masses, there must be  four light Nambu-Goldstone bosons,  but in Nature we observe only three pions (of  $SU_A(2)$ breaking).
Where is the fourth NG boson?     A possible fourth NG boson, $\eta$, has  actually mass 
\be      m_{\eta}  \gg    m_\pi  \; 
\ee
(the $U(1)$ problem).

The basic answer is given by 't Hooft:   the axial $U(1)$ current  satisfies an anomalous divergence equation
\be    \partial^{\mu}  J^{(A)}_{\mu} =     N_f\,   \frac{g^2}{32 \pi^2}  F_{\mu\nu} {\tilde F}^{\mu \nu}\;, \qquad   {\rm with} \quad  \int d^4x  \,  \frac{g^2}{32 \pi^2}  F_{\mu\nu} {\tilde F}^{\mu \nu} = {\mathbbm Z}\;, 
\ee
where ${\mathbbm Z}$ is the integer instanton number.

An efficient way to represent the strong anomaly effects is that of writing a low-energy effective action, containing the term reproducing the  correct $U_A(1)$  symmetry breaking
\cite{Witten:1979vv}-\cite{Nath},
\be {L} =  {L}_0 +    {\hat  L}\;,   \qquad    {\hat  L}=    \frac{i}{2} \, q(x) \, \log   \det U/U^{\dagger}  +   \frac{N}{a_0  F_{\pi}^2 }   q(x)^2\;;  \label{effL}
\ee 
where   $L_0$ is the standard chiral Lagrangian describing the massless  pions, 
\be      U =   U_0 \,  e^{ i \pi^a(x)  t^a    / F_{\pi}}    \label{pion}  
\ee
and 
\be    q(x) =   \frac{g^2}{32 \pi^2}  F_{\mu\nu} {\tilde F}^{\mu \nu}\;.  \label{topodens}
\ee
Such an effective Lagrangian correctly reproduces the anomalous  $U_A(1)$ variation, 
\be  \Delta S =    2 N_f \alpha \int d^4x  \frac{g^2}{32 \pi^2}  F_{\mu\nu} {\tilde F}^{\mu \nu}\;, \qquad \psi_{L,R}   \to   e^{\pm i \alpha} \psi_{L,R}\;,
\ee
and would give a mass $\propto \Lambda$ to $m_{\eta}$.  
}

The question is:    does a (multi-valued)  logarithmic function   make sense as an effective Lagrangian?

The answer is: yes, it does, 
 if  $U$  acquires a nonvanishing VEV, and   if (\ref{effL})  is regarded as a function of the  pion field $\pi(x)$, i.e., as an expansion, 
 \be      U =   \brc  U \ckt    e^{ i \pi^a(x)  t^a    / F_{\pi}}   =    \brc  U \ckt    (1 +   i \pi^a(x)  t^a    / F_{\pi} +\ldots )\;.
\ee
Now, the idea is to invert the logic,  and say that  an effective action (\ref{effL})   with the logarithmic,  anomaly term,  {\it requires}  that the chiral composite 
$U\sim    {\bar {\psi}_R} \psi_L $  to get a nonvanishing VEV,   (\ref{VEV}).  In other words, {\it    the strong anomaly  implies  the spontaneous symmetry breaking }  of the 
chiral   $SU_L(2) \times SU_R(2) \times U_V(1) \times U_A(1)$ symmetry  to the diagonal, vector subgroup,   $SU_V(2)\times U_V(1)$.

We now apply the same idea to chiral gauge theories,   where  the form of the strong anomaly is known but not the dynamical, infrared phase. For concreteness, let us take the 
``$\chi \eta$ model"   (see (\ref{GGmodels})),  with fermions, 
\be       \yng(1,1) \oplus   (N-4) \,{\bar   {{\yng(1)}}}\,.     \label{chieta}    
      \ee
The form of the strong anomaly  is known: 
\be         \frac{i}{2}    q(x)      \log      (\chi\eta)^{N-4} \chi\chi  +{\rm h.c.}\;,       \label{stronganomalychieta}    \ee    
($q(x)$ is the topological density defined in (\ref{topodens}))  where 
\bea    && (\chi\eta)^{N-4} \chi\chi  \equiv     \epsilon_{i_1 i_2 \ldots i_N}   \epsilon_{m_1 m_2 \ldots m_{N-4}}  \,  (\chi \eta)^{i_1 m_1}    (\chi \eta)^{i_2 m_2} \ldots  (\chi \eta)^{i_{N-4} m_{N-4}}    \chi^{i_{N-3}i_{N-2} }  \chi^{i_{N-1}i_{N} }\;. \nn \\  \label{zeromodes}
  \eea
  Now such a strong-anomaly effective action implies  
  \be   \brc \chi\eta\ckt\ne 0\;, \quad     \brc \chi\chi\ckt\ne 0\;: 
  \ee
i.e., the system is  in the dynamical Higgs phase   \cite{BKL5,Vene}. 

Note that the confining, chirally symmetric phase, with no condensates, and the massless ``baryons",  ${\cal B} \sim    \chi\eta\eta$,
as  the only  infrared degrees of freedom, fails the ``matching" of the strong anomaly.  The strong anomaly effective action  above cannot be written in  terms of  
${\cal B}'s$,  as the  fermion zero mode counting (in the instanton background) fails.  

It can be shown that the strong-anomaly consideration favors the dynamical Higgs phase, against the confining symmetric phase with no condensate formation, 
in all BY and GG models \cite{BKL5}, in agreement with the indication coming from the studies of the generalized anomaly matching (especially, the ${\mathbbm Z}_2$ anomalies) 
studied in Sec.~\ref{Z2anomaly}.

Such an agreement is not really a coincidence.  Both are consequences of taking into account appropriately the effects of the strong anomalies  (i.e., topologically nontrivial, nonperturbative effects of the strong $SU(N)$ gauge dynamics).

\subsection{Dynamical Abelianization  \label{DynAb} } 

Another interesting result found concerns the ``$\psi\chi\eta$"  (and some other)  model,   with fermions,
\be   \psi^{\{ij\}}\;, \qquad  \chi_{[ij]}\;, \qquad    \eta_i^A\;,\qquad  A=1,2,\ldots 8\;,       \label{psichieta}
\ee
or
\be     \yng(2) \oplus  {\bar  {\yng(1,1)}}  \oplus   8  \times  {\bar  {\yng(1)}}\;.  
\ee
It  is  asymptotically free, the first coefficient of the beta function being,
\be   b_0  = \frac 13\left[ 11N-  (N+2) -(  N-2) -8  \right] =  \frac{ 9 N- 8 }{3}\;.      \label{beta0}
\ee
The model has a global $SU(8)$ symmetry as well as two $U(1)$  symmetries,  
\be   \tilde U(1): \qquad   \psi\to  e^{2i\alpha} \psi\;, \quad \chi \to   e^{ -2i\alpha} \chi\;, \quad \eta \to   e^{ -i\alpha} \eta\;, \label{def:tildeU1} \ee
and
\be U(1)_{\psi\chi}: \qquad   \psi\to  e^{i  \frac{N-2}{N^*} \beta} \psi\;, \qquad \chi \to   e^{- i \frac{N+2}{N^*}\beta} \chi\;,   \qquad \eta \to   \eta \;, \label{def:U1psichi}\ee  
where 
\be N^*=GCD(N+2, N-2)\quad \text{and}\quad  \alpha,\;\beta\in (0, 2\pi)\;.\ee
The problem is to understand how these symmetries  and what kind of phase,   are realized at low energies.

We adopt  the following strategy:   our initial investigation of this model   \cite{BKS,BK} has shown that the conventional 't Hooft anomaly matching algorithm  allows, among few others,  the dynamical Abelanization.   The idea is to study whether  the consequences of this dynamical assumption expected in the infrared are consistent with the indications of possible  
generalized 't Hooft anomalies, in the ultraviolet. These provide  stronger constraints than the conventional anomaly-matching algorithm.

We assume that bifermion condensates  in the adjoint representation
\be   
\langle  \psi^{ik} \chi_{kj}   \rangle  = \Lambda^3   \left(\begin{array}{ccc}
      c_1  &  &  \\     & \ddots   &   \\   &  &    c_{N}
  \end{array}\right)^i_j   \;, \qquad  \brc  \psi^{ij}  \eta_j^A  \ckt    = 0
  \;,   \label{psichicond}  
\ee
\be         c_n    \in {\mathbbm C}\;,     \qquad     \sum_n c_n =0\;, \qquad       c_m - c_n  \ne 0\;, \ \     m \ne n  \;,  \label{psichicondBis}  
\ee
(with no other particular  relations among $c_j$'s)   
form  in the infrared,   and 
 induce  the symmetry breaking  
\be    
SU(N)    \to    U(1)^{N-1} \;.      \label{SUNbreaking}    \ee
   a phenomenon well-known in the ${\cal N}=2$ supersymmetric gauge theories  \footnote{In contrast to ${\cal N}=2$ susy theories,   here the scalar in the adjoint representation   $\phi \sim \psi\chi$    appears as a dynamical, composite field.  
     }. 
     In \cite{BKL5}  a detailed study was made of all aspects of symmetry realization in low energies, including the effects of the strong anomalies. 
In particular,   analysis of    the consistency with the gauging (and the associated 't Hooft anomalies)  of the  1-form   color-flavor locked ${\mathbbm Z}_N$  symmetry  
(see Table~\ref{symmetries}  below)  
give convincing 
evidence  that the assumption of the dynamical Abelianization is a correct one.  (See also \cite{Sheu:2022odl}).
\begin{table}[h]
  \centering 
  \begin{tabular}{|c |c|c| c|c|c|c|c|c| }
\hline
$ \phantom{{{   {  {\yng(1)}}}}} \!  \! \! \! \! \!\!\!$ &                $  {\tilde U}(1)   $   &     $  U(1)_{\psi\chi}   $   &    $({\mathbbm Z}_{N+2})_{\psi}$ &  $({\mathbbm Z}_{N-2})_{\chi}$       &    $SU(8)_{\eta}$      &   ${\mathbbm Z}_{N^*}$     &     $  {\mathbbm Z_{4/N^*}}$
     \\   
 \hline 
 Mixed Anomalies &     \checkmark   &  X  &   X  &   X  &         \checkmark    &      \checkmark   &      \checkmark     \\
 \hline
   Dyn. Abel.     &     \checkmark   &  X &    X &   X  &      \checkmark    &    \checkmark    &     \checkmark    \\
 \hline
\end{tabular}  
\caption{\footnotesize   Dynamical Abelianization postulate of the present work is confronted with the implications of the mixed anomalies.
      $\checkmark $   for a conserved symmetry,  X     for a broken symmetry.  The discrete  ${\mathbbm Z}_{N^*}$   symmetry is defined in   \cite{BKL5}. 
    }\label{symmetries}
\end{table}

\subsection{More general dynamical symmetry breaking patterns}

There are many systems other than the $\psi\chi\eta$ model (\ref{psichieta}) in which a bifermion condensate in  the adjoint representation can  form.  An interesting class of models are  those 
(see  (\ref{tensors}))   with fermions  in second-rank tensor representations,    
\be     \tfrac{N- 4}{k}   \,  \psi^{\{ij\}}  \oplus   \tfrac{N+4}{k} \,  {\bar  \chi_{[ij]} }\;.    \label{tensor}   \ee 
  It is plausible that a bifermion condensate
\be   \brc  \psi \chi \ckt  \ne 0 
\ee
forms,  but   it may not necessarily imply the dynamical Abelianization  (as we assumed for the  $\psi\chi\eta$ model).   The condensate can have the form  \cite{BKLproc}, 
\be      \brc  \psi \chi \ckt  = {\rm diag.}\,   ( c_1 {\mathbf 1}_{n_1}, c_2 {\mathbf 1}_{n_2},  \ldots   )\;, \qquad   \sum_i  c_i  n_i   = 0\;,
\ee
leading to a dynamical color symmetry breaking, 
\be     SU(N)   \to    SU(n_1)\times SU(n_2) \times \cdots   \prod_k  U_k(1) \;.      \ee
In particular, we wish to know whether some of these nonAbelian subgroups can be infrared-free,  i.e.,  can survive in the infrared.    
Such a question is relevant because the  standard model of the fundamental interactions is based on a gauge theory having precisely the gauge group of this type.  

\subsubsection{$N=5$, $k=1$ model}

For instance,  consider the model (\ref{tensor}) with  $N=5$, $k=1$,   with fermions
\be      \yng(2) \oplus   9\cdot  {\bar {\yng(1,1)}}\;.    
\ee
A possible pattern is 
\be    G_c=  SU(5) \to     SU(3)\times SU(2) \times U(1)\;. 
\ee
The global symmetry is broken as  
\be  G_F = SU(9) \times U_0(1)     \to   SU(8)    \times U_0(1)^{\prime}   \;.
\ee
$ U_0(1)^{\prime} $  is the  combination  
\be     e^{ i \alpha   \left(\begin{array}{cc}  {\mathbf 1}_8/8  & 0 \\0 & -1  \end{array}\right)}\in SU(9)\; \quad {\rm with}    \; \quad    e^{ i \beta  Q_0} \;, 
\ee
that is,  with
\be     -  \alpha + \beta (  \frac{9}{N+2} -  \frac{1}{N-2} ) =0\;, \qquad \alpha= 4\,  \frac{ 2N - 5}{N^2-4}\, \beta \;.
\ee

The fermions, decomposed in the representations of the unbroken subgroup are  shown in Table~\ref{Simplest}. 
\begin{table}[h!t]
  \centering 
  \begin{tabular}{|c|c|c |c|c|c|c| }
\hline
$ \phantom{{{   {  {\yng(1)}}}}}\!  \! \! \! \! \!\!\!$   & fields      &  $ SU(3)$     &   $ SU(2)  $     &   $U(1)$   &     $SU(9)$  &   $U_0(1)$        \\
 \hline
  \phantom{\huge i}$ \! \!\!\!\!$  {\rm UV}
  & $\psi^{ij} $     &    $  \yng(2) $    & $  (\cdot )$   &   $4$       &    $  (\cdot )$    &     $ \frac{9}{N+2} $     \\
  &  $\psi^{iJ} $      &    $  \yng(1)  $    & $ \yng(1) $   &  $-1$         &    $  (\cdot )$  &    $ \frac{9}{N+2} $         \\
 &  $ \psi^{JK}$      &   $  (\cdot )$      &   $  \yng(2)  $  &   $-6$      &    $  (\cdot )$    &      $ \frac{9}{N+2} $         \\
  &  $ \chi_{ij}^A$      &   $ {\bar  {\yng(1,1)}} = \yng(1)   $     &  $  (\cdot )$   &   $-4$      &    $  \yng(1)  $   &      $- \frac{1}{N-2} $    \\
   &  $ \chi_{iJ}^A$      &   $ {\bar {\yng(1)} }  $     &   $  \yng(1)  $  &     $1$       &  $  \yng(1)  $     &      $- \frac{1}{N-2} $     \\
    &  $ \chi_{JK}^A$      &   $  (\cdot )$      &   $  (\cdot )$   &   $6$      &     $  \yng(1)  $    &        $- \frac{1}{N-2} $     \\
\hline
\end{tabular}  
  \caption{\footnotesize     $\psi\chi$ model, $N=5$, $k=1$.    $A=1,2,\ldots, 9$, 
  $i,j=1,2,3$;   $J,K=4,5$.
     }
\label{Simplest}
\end{table}
The fermions  participating in the condensate, $\psi^{iJ}  \chi_{J i }^9$,    become massive,  and leave the massless fermions in Table~\ref{5132}.  
\begin{table}[h!t]
  \centering 
  \begin{tabular}{|c|c|c |c|c| c| c| }
\hline
$ \phantom{{{   {  {\yng(1)}}}}}\!  \! \! \! \! \!\!\!$   & fields      &  $ SU(3)$     &   $ SU(2)  $     &   $U(1)$     &    $SU(8)$   &  $U_0(1)^{\prime}  $    \\
 \hline
  \phantom{\huge i}$ \! \!\!\!\!$  {\rm IR}
  & $\psi^{ij} $     &    $  \yng(2) $    & $  (\cdot )$   &   $4$      &    $  (\cdot )$      &      $ 2(N-2)$    \\
 &  $ \psi^{JK}$      &   $  (\cdot )$      &   $  \yng(2)  $  &   $-6$       &  $  (\cdot )$         &        $ 2 (N-2)  $      \\
  &  $ \chi_{ij}^9$      &   $ {\bar  {\yng(1,1)}} = \yng(1)   $     &  $  (\cdot )$   &   $-4$      &     $  (\cdot )$   &   $ -  2 (N-2)  $       \\
    &  $ \chi_{ij}^B$      &   $ {\bar  {\yng(1,1)}} = \yng(1)   $     &  $  (\cdot )$   &   $-4$      &   $ \yng(1) $    &     $-1 $      \\
     &  $ \chi_{iJ}^{B}$      &   $ {\bar {\yng(1)} }  $     &   $  \yng(1)  $  &     $1$      &  $ \yng(1) $    &        $ - 1 $    \\
    &  $ \chi_{JK}^9$      &   $  (\cdot )$      &   $  (\cdot )$   &   $6$      &    $  (\cdot )$     &          $ - 2(N-2)  $      \\
        &  $ \chi_{JK}^B$      &   $  (\cdot )$      &   $  (\cdot )$   &   $6$      &  $ \yng(1) $      &      $ - 1 $    \\
        \hline
        &  $\phi^B \sim  \Re   (\psi^{iJ}  \chi_{J i }^B) $   &        $  (\cdot )$      &   $  (\cdot )$   &    $  (\cdot )$     &  $ \yng(1) $      &      $2N-5$         \\
         &  $\pi^B  \sim  \Im   (\psi^{iJ}  \chi_{J i }^B)   $   &        $  (\cdot )$      &   $  (\cdot )$   &    $  (\cdot )$     &  $ \yng(1) $      &     $  2N-5 $      \\
          &  $\pi^9\sim  \Im   (\psi^{iJ}  \chi_{J i }^9)$   &        $  (\cdot )$      &   $  (\cdot )$   &    $  (\cdot )$     &   $  (\cdot )$      &     $0$    \\
        
\hline
\end{tabular}  
  \caption{\footnotesize   Massless fermions and Goldstone bosons in the infrared, in  $N=5$, $k=1$,  $\psi\chi$ model. $B=1,2,\ldots, 8$.     }
\label{5132}
\end{table}
The ``low-energy"  $SU(3)$  group turns out to be asymptotically free  
(evolve to stronger interactions in the IR,  leading to further dynamical symmetry breaking or confinement), whereas  the $SU(2)$ is infrared free:  
 are  still asymptotically free: 
\be    \beta_{SU(3)} =    11\cdot 3 -  5 -  9   \cdot  1   -  8   \cdot  1  \cdot 2   >0\;, \ee
\be     \beta_{SU(2)} =    11\cdot 2 -  4  -   8\cdot 3\cdot 1   <    0\;. 
\ee
This could be interesting, in principle. After all,   the standard $ SU(3)\times SU(2) \times  U(1) $ model   contains also an asymptotically-free  sub-gauge group $SU(3)$   
and the infrared-free $SU(2)\times U(1)$ part.  Unfortunately the matter content is not quite the same, although one can notice intriguing similarities.

\subsubsection{Infrared nonAbelian gauge groups    \label{generalDSB}} 
More generally,  we  ask whether or not  the subgroup  
\beq
SU(N) \to SU(n) \times \dots
\eeq
in (\ref{generalDSB})  can remain infrared-free  (weakly coupled) at low energies, in a given model.   By decomposing the fermions  (\ref{tensor})  as direct sums of the irreps of  the $SU(n)$ subgroup,  
the (first coefficient the)  beta function of $SU(n)$ can be seen to be 
\bea
\beta(SU(n)) &=& 11 \cdot n - \frac{N-4}{k}\cdot (n+2) - \frac{N+4}{k}\cdot (n-2) - \frac{8}{k}\cdot(N-n)\cdot 1   \nonumber \\  &=& \frac{16+8n+11 k n -8 N -2 n N}{k}\;.  
\eea
The change of sign happens at 
\beq
n^*= \frac{8N -16}{8 + 11k -2N}
\eeq
so the integer part of $n^*$, $[n^*]$ is the biggest $n$ that is IR free. Clearly if $[n^*]=1$ there are no non-abelian IR free (gauge) symmetry breaking patterns. Let us discuss a few  cases. 

\begin{itemize}
\item $k=1$,  $N=5$,  $
n^*=   \frac{8}{3}
$,    so $[n^*]=2$. Indeed 
\beq
\beta(SU(2)) =-6\;.
\eeq
The possible IR free breaking is 
\beq
 SU(5)   \to   SU(2)\times SU(2) \times  U(1)^2\;.
\eeq
\item $k=1$,  $N=6$,   $
n^*=    \frac{32}{7}
$,    so $[n^*]=4$, and 
\beq
\beta(SU(4)) =-4\;. 
\eeq
A possible IR free breaking mode  is (see Fig.~\ref{ZNfig6142})
\beq
 SU(6)   \to   SU(4)\times SU(2) \times  U(1)  \;.
\label{b16}
\eeq

In the following,  the massless fermions which remain in the infrared  are shown in quiver graphs, instead of using tables  such as Table~\ref{5132}. 
Circles with a number inside $n$ represent a gauge group $SU(n)$, squares with a number $m$ represent a global symmetry $SU(m)$ group, fermions are lines connecting the groups, arrows on the line indicate if is fundamental (ingoing) or antifundamental (outgoing),  lines omitted for $SU(2)$; little ``o'' or ``x'' within a line indicates if they belong to symmetric or anti-symmetric tensor representation.

\begin{figure}[h!]
\begin{center}
\includegraphics[width=1.5in]{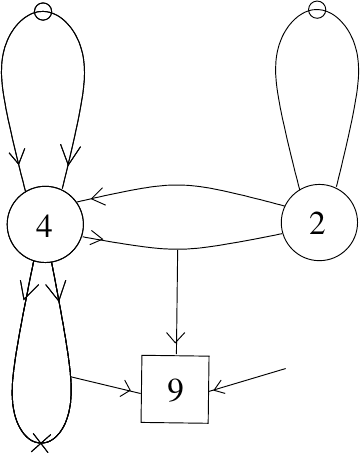}
\caption{\footnotesize Diagram corresponding to   (\ref{b16}). }
\label{ZNfig6142}
\end{center}
\end{figure}

Another possible IR free breaking mode  is  (see Fig.~\ref{ZNfig6133})
\beq
 SU(6)   \to   SU(3)\times SU(3) \times  U(1) \;.
\label{b26}
\eeq
\begin{figure}[h!]
\begin{center}
\includegraphics[width=1.5in]{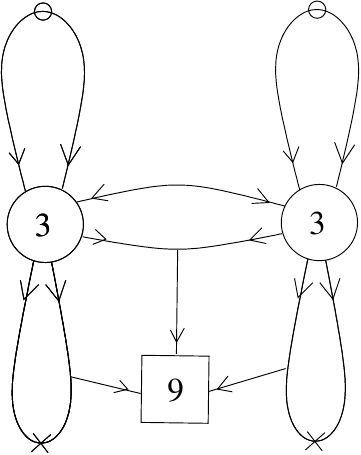}
\caption{\footnotesize Quiver diagram corresponding to  (\ref{b26}) }
\label{ZNfig6133}
\end{center}
\end{figure}
\end{itemize}
\begin{itemize}
\item $k=2$,   $N=6$,  $n^*=    \frac{16}{9}$.    Since $[n^*]=1$  there are no non-abelian IR free symmetry breaking patterns. Abelianization is the only IR-free possibility.
\item $k=2$,  $N=8$,  $n^*=  3+ \frac{3}{7}$   and  
\beq
\beta(SU(3)) =-3\;.  
\eeq
Possible IR free breaking patterns are
\beq
 SU(8)   \to   SU(3)\times SU(3)\times SU(2) \times  U(1)^2
\label{b18}
\eeq
or
\begin{figure}[h!]
\begin{center}
\includegraphics[width=2.5in]{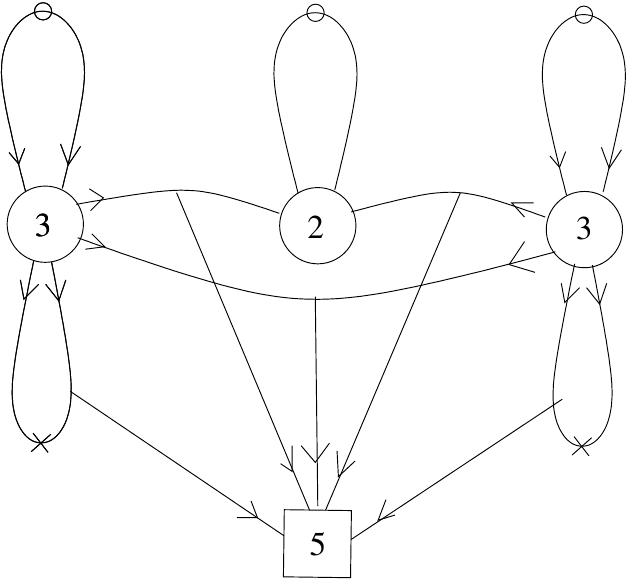}
\caption{\footnotesize Diagram corresponding to (\ref{b18}) }
\label{ZNfig82332}
\end{center}
\end{figure}
\beq
 SU(8)   \to   SU(2)^4\times     U(1)^3\;.
\eeq
\item $k=2$,   $N=10$,  $
n^*=  \frac{32}{5}$,       
 so $[n^*]=6$, and indeed   
\beq
\beta(SU(6)) =-2\;.
\eeq
Possible IR free breaking modes  are
\beq
 SU(10)   \to   SU(6)\times SU(4) \times  U(1) 
\label{b210}
\eeq
and
\begin{figure}[h!]
\begin{center}
\includegraphics[width=1.8in]{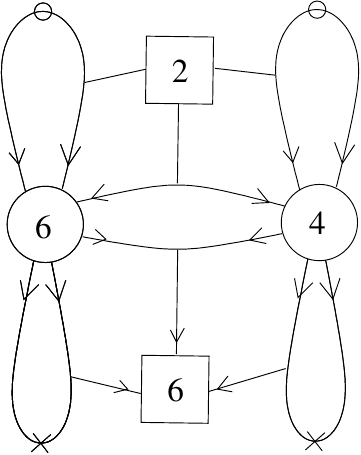}
\caption{\footnotesize Diagram corresponding to (\ref{b210}) }
\label{ZNfig}
\end{center}
\end{figure}
\beq
 SU(10)   \to   SU(5)\times SU(5) \times  U(1) \;,
\eeq

\end{itemize}

and so on.

  It is left for further study to understand which is the correct phase of each model.

\section{Old and New criteria for confinement and other phases   \label{Oldcriteria}} 

These efforts to understand the dynamics and phases of chiral gauge theories in four dimensions  reviewed above,  remind us of the well debated confinement (or Higgs) criteria, 
in particular  in the context of pure Yang-Mills theory or of QCD,    and  urge us to revisit these ideas with more critical eyes.

There are three well-known ``confinement criteria".  (A):  The original idea that a colored particle cannot  be freely propagating: they are confined inside a color-singlet composite object. 
(B):  Criteria which use Wilson loop or Polyakov loop;  and (C): The dual superconductivity idea by 't Hooft.    As will be seen,   each of them has some issues.

\begin{description}

  \item[(A)]   The original color-confinement idea that colored particles (e.g., quarks) cannot be freely propagating, and permanently confined inside a color-singlet 
composite states  (e.g., hadrons). 

 This concept, which seems to be well-defined,    appears to be somewhat problematic, 
when applied to some chiral gauge theories. Namely, gauge non-invariant operators or states could well 
 be  gauge-invariant ones, just written in a particular gauge.  
 
  A noteworthy and well known example is the Weinberg-Salam  $SU(2)\times U(1)$ theory.  It is usually stated that 
  the $SU(2)\times U(1)$  gauge group is spontaneously broken by the Higgs   VEV
  \be     \brc \phi\ckt =    \left(\begin{array}{c}  {\rm v}   \\0\end{array}\right)\;, \qquad  {\rm v}=  256 \, {\rm GeV}\;, \label{Higgs} 
  \ee
  and the neutrino and the lefthanded electron are the upper and lower components of the fermion lefthanded doublet, 
  \be    \psi_L=      \left(\begin{array}{c}   \nu   \\   e_L    \end{array}\right)\,.  \label{leptons} 
  \ee
  A more appropriate way to think about these is that (\ref{Higgs}) really means that 
   \be     \brc   \sum_{i=2}^2    \phi_i^{\dagger}  \phi^i   \ckt  \ne  0  \;:    \label{Higgsvero} 
  \ee
  and that the neutrino and electron are described by  gauge invariant composite fields
  \be   \nu \sim  \phi^{\dagger}_i \psi_L^i \;, \qquad   e_L  \sim  \epsilon_{ij}  \phi^i  \psi_L^j \;.
  \ee
The familiar expressions  (\ref{Higgs}) and (\ref{leptons}) are just formulae valid in the particular (and arbitrary) gauge chosen, (\ref{Higgs}).    
  
    Does it mean that there are no distinctions between the confinement  and Higgs phases?    The answer is: there are.  The weakly coupled, $SU(2)\times U(1)$ theory in broken,  Higgs phase,
    cannot be understood as a strongly coupled $SU(2)\times U(1)$ theory in confinement phase \cite{Abbott}. 
      Analogous remarks can be made in certain classes of chiral gauge theories studied above.  In particular, in BY and GG models which  are  likely to be in a dynamical Higgs phase,  there is no difficulty in rewriting the condensate such as  (\ref{DHiggs}) or the NG boson of the system,  in  a gauge-invariant fashion.   Nevertheless, the dynamical Higgs phase 
     is clearly  distinct from the putative, confining, symmetric phase  \cite{BKL2,BKL4,BKLReview}.

  \item[(B)]   Another well-known criterion uses  the Wilson loop  
  \be     W_{\gamma} =  \Tr  \,\{  {\cal P} e^{i \oint_{\gamma}  A_{\mu} dx^{\mu}  }   \}   \;,
  \ee
  or  the Polyakov loop in Euclidian spacetime,
  \be   P({\bf r}) = \frac{1}{N}  \Tr \{  {\cal T} e^{ i \int_0^{\beta}   d\tau    A_0({\bf r}, \tau) }  \}  \;, 
  \ee
  where ${\cal P}$ and ${\cal T}$ represent the path-ordered or time-ordered exponentials. 
  Wilson's criterion is 
  \be     \lim_{\gamma \to \infty}    \brc    W_{\gamma}  \ckt =  
   \begin{cases}
    e^{- A}    & \text{confinement phase}, \\
     e^{-L}    & \text{Higgs phase}:     \label{Wilson} 
\end{cases}
  \ee
  i.e., the area law - a linearly rising potential between two test charges -  indicates confinement phase. 
  
   The Yang-Mills theory is invariant under  the center symmetry transformation  of the Polyakov loop, 
  \be     P({\bf r})  \to     {\mathbbm Z}_N \,    P({\bf r}) \;. 
  \ee
   The unbroken center symmetry 
   \be    \lim_{\beta \to \infty}   |  \brc  P({\bf r})  \ckt  |  = 0\;,   \label{Polyakov}  
   \ee
   can be used as a criterion of confinement phase  (an infinite free energy for an isolated quark).

   The lattice simulation indicates that  the $SU(N)$ Yang-Mills theory is indeed in confinement phase, according to these criteria.

   The problem with this criterion  is that there is nothing to confine in pure YM theory (!).   As soon as massless quarks are introduced,  center symmetry is broken.  Also, 
   the confining string between two test charges is broken by the spontaneous production of quark-antiquark pairs  from the vacuum (vacuum polarization),  and the area law is lost.  
   The perimeter law ensues.  Thus neither (\ref{Wilson}) nor (\ref{Polyakov})  can be used  to  discriminate the infrared phases  (Higgs or confining) of quantum chromodynamics (QCD) with massless quarks.

    \item[(C)]   Confinement as a dual Meissner effect  ('t Hooft).    
    By considering the subgroup of the color $SU(3)$,
    \be  U(1)^2  \subset SU(3)   \;,
    \ee
    the charges of a particle (elementary or solitonic)  can take electric and magnetic quantum numbers
    \be         (n_1, n_2; m_1,m_2)\;,  \qquad   n_i, m_i \in  {\mathbbm Z}  \;, \quad i=1,2\;.  \label{charges} 
    \ee
   The corresponding $U_i(1)$ electric and magnetic coupling strengths  are  
    \be      n_1 e_1\;,    m_1  g_1\;,    n_2 e_2\;,    m_2 g_2\;,  
    \ee
    where the elementary electric and magnetic charges obey Dirac's quantization condition, 
    \be    e_1 \, g_1  \in {\mathbbm Z}/ 2\;, \qquad   e_2 \, g_2  \in {\mathbbm Z}/ 2\;.
    \ee
      Now define the ``Dirac unit" between two particles  
      \be   {\cal D}^{1,2}  \equiv    \sum_{i=1}^2   ( n_i^{(1)}  m_i^{(2)}  -   n_i^{(2)} m_i^{(1)}  )\;.  
      \ee
Then  't Hooft's criterion \cite{tHooftConf}  is the following. If the field of particle $1$ with charges (\ref{charges})  condenses  
\be   \brc  M^{(1)} \ckt  \ne 0\;,  
\ee
then all  particles  2  carrying charges with nonvanishing Dirac unit with respect to particle  1,
\be       {\cal D}^{1,2}  \ne   0\;,    \quad {\rm  Mod} (3)\;, 
\ee
are confined.

  For instance,  the  condensation of the magnetic monopole of   $U_1(1)$,
\be     \brc  M_{0,0;  1,0} \ckt  \ne 0\;,  
\ee
implies that the quark with charge   $Q_{(1,0;0,0)}$ is confined  (dual Meissner effect).

The criterion (C) is also problematic.  In formulating the confinement criterion in terms of  the  $U(1)^2  \subset SU(3)$  charges, 
{\it   one has made an implicit, dynamical assumption}  that these Abelian  (electric, magnetic or dionic)  degrees of freedom describe the physics in the infrared. In other words, one assumes
dynamical Abelianization,   analogous to what happens in ${\cal N}=2$  supersymmetric gauge theories or  in the  chiral  $\psi\chi\eta$  model  we discussed in  Sec.~\ref{DynAb}.  
However, in the standard QCD   there are no  elementary or  (plausible) composite scalar fields in the adjoint representation \footnote{Actually there are  bifermion $\psi_L  {\bar {\psi_R}} $  or   bi-gluon  
$G_{\mu\nu} G^{\mu \nu}$  composites which may act as scalar fields in the adjoint representation. However the corresponding composite scalars in the color-singlet representation  are in a much more strongly attractive channels, and indeed  those are believed to form condensates in the real-world QCD.  
},  in contrast to these other systems.  In such a situation dynamical Abelianization of the
system is unlikely.  Besides, there are no phenomenological indications in favor of an Abelian  $U(1)^2$  infrared effective theory for QCD. 

It is possible that  confinement in QCD  is explained by  a dual superconductor mechanism,  but without Abelianization.  But it means that the infrared physics involves nonAbelian monopoles
 and their quantum dynamics,  a notoriously subtle problem.    
See e.g.  \cite{75}  for a review and for references to earlier literature.

Quantum-mechanical properties of Abelian or nonAbelian monopoles and dyons, their dynamics and 
 their possible roles in confinement and symmetry breaking,  have, on the other hand,    been largely clarified by the ground-breaking discovery of the exact Seiberg-Witten solutions of ${\cal N}=2$ supersymmetric gauge theories \cite{SW1}-\cite{Tachikawa}. Unfortunately,   it turns out
 that it is rather difficult to make reliable predictions about ordinary (i.e.,  nonsupersymmetric) theories, by using the knowledge gained in the context of ${\cal N}=2$ 
(or ${\cal N}=1$) supersymmetric theories.    In general, one expects various phase transitions, when the ${\cal N}=2$  or ${\cal N}=1$  susy breaking terms are added
in the action, and are tuned to be larger than the dynamical scales $\Lambda_{{\cal N}=2}$  or  $\Lambda_{{\cal N}=1}$ of supersymmetric theories.    

Under these circumstances,  the best thing one can do could be to try to learn the kind of physics phenomenon which is dynamically realizable, and which   
could be underlining the confinement in QCD, rather than attempting to deform the  ${\cal N}=2$  or  ${\cal N}=1$ QCD in some concrete manner  (see  \cite{Csaki1,Csaki2,Csaki3}  for recent efforts), hoping to get something which looks  similar to the standard QCD.  

We shall indeed take the first attitude,  and discuss below an idea, realized in susy gauge theories, and might be underlying the physics of real-world QCD, that confinement is a deformation of the RG flow towards a nonAbelian strongly-coupled 
conformal fixed point  (Sec.~\ref{lesson}). 

But before that, let us discuss new simple criteria for "color confinement", Higgs phase, etc.

\end{description}

\subsection{New criteria (tentative)  for different phases \label{Newcriteria}}

The difficulties in the familiar  "confinement criteria"  (A) - (C)  reviewed above indeed tempt us to propose the following, new criteria  for different phases of 
  chiral  or vectorlike $SU(N)$ gauge theories in four dimensions.

  Let us however keep note of a lesson, first of all,  that  the studies in different classes of chiral  gauge theories discussed in Sec.~\ref{newmore},
all based on the same $SU(N)$ Yang-Mills theory,  taught us.   It is the fact that    the infrared dynamics (and the phase) of  an asymptotically-free  $SU(N)$ gauge 
theory     {\it      is  not determined by that of the pure $SU(N)$ Yang-Mills theory on which it is based, but  by the dynamics involving the massless matter fields,} which in turn depend critically on 
their representation.  Thus an argument that  a chiral gauge theory (e.g., the BY model)  should necessarily confine, as the $SU(N)$ Yang-Mills theory is in  confinement phase,   is logically unfounded.

 Our proposal is that the different phases are characterized by the number and types of various colored Nambu-Goldstone bosons the system produces.  
 They may be generated  by the condensation of either elementary or composite scalar fields.  
We exclude below those systems which are infrared-free $SU(N)$ theory (many fermions are present so that the interactions are weak at low energies.  $SU(N)$ gauge bosons survive in the IR  as asymptotic states    - Coulomb phase. We exclude also  
 those which flow into conformal fixed-point theories (however, this last class of models may have a subtle and important relation to confinement, see below).

\begin{description}
  
  \item[(a)]   The system is in the confinement phase if it produces no colored NG bosons.
  
  This is the case for YM,  QCD, supersymmetric QCD and  susy YM.   Note that the  phase of the standard QCD with massless fermions is classified as confinement according to this new criterion,
   whereas   the old criterion  (B)  fails.   
  
  \item[(b)]  (Dynamical) Higgs phase, when the system produces $N^2-1$ colored NG bosons.
  
  This occurs (likely)  in the chiral BY and GG models studied in  \cite{BKL1}-\cite{BKLReview}, and in the Glashow-Weinberg-Salam electroweak theory.

  \item[(c)] Dynamical Abelianization (or Coulomb phase)  occurs when there are $N(N-1)$ colored NG bosons.
  
  This was shown likely to be the correct phase of the $\psi\chi\eta$ model \cite{BKLDA,Sheu:2022odl}, and is known to be realized in most ${\cal N}=2$ supersymmetric gauge theories \cite{SW1}-\cite{Tachikawa}.

  \item[(d)]   Other groups other than  (a)-(c) above,   of  colored NG bosons. 
  
  The system could flow into infrared effective theory containing some residual (infrared-free) nonAbelian gauge dynamics.  The first attempts investigating these possibilities in chiral theories are reported in 
  Sec.~\ref{generalDSB}  above.    In the context of ${\cal N}=2$ supersymmetric gauge theories,  this type of infrared-fixed-point theories are well-known (e.g., the $r$-vacua of ${\cal N}=2$  SQCD \cite{APS,CKM}.)

\end{description}

These new criteria clarify to some extent the ideas about possible different phases occurring in various types of strongly-coupled gauge theories in four dimensions, chiral or vectorlike. 
Still,  confinement in QCD with massless quarks certainly requires a better explanation than the earlier criteria  (A)-(C) reviewed in Sec.~\ref{Oldcriteria}, and 
a more detailed understanding of the mechanism than  (a).

Below, we discuss an idea on the confinement in QCD, which might sound somewhat extraordinary, but is indeed realized in some softly-broken supersymmetric theories.

\section{A lesson from supersymmetric gauge theories  \label{lesson}}

As already said, it remains an unsolved problem to make reliable  predictions about the dynamics of  any  specific (non-supersymmetric) gauge theory such as QCD, 
based on the knowledge about supersymmetric gauge theories.

But the  result of studies on supersymmetric gauge theories during the last 50 years does teach us what sort of dynamical phenomena are possible in strongly-coupled nonAbelian gauge theories, how they depend on the gauge group and on the massless matter 
contents, and how they work concretely. {\it   It has given solid understanding of the nonperturbative effects involving  instantons, magnetic monopoles, dualities and interacting (super) conformal infrared fixed points  (SCFT). }

From the point of view of the  renormalization-group flow,  confinement can be understood as a deformation (deviation)  by some relevant operator (which might be present already in the UV theory or 
produced dynamically),   from a trajectory leading to an infrared fixed  point.  See Fig.~\ref{figRGflow}. 
The relevant  infrared fixed point theory might be Abelian (Abelianization),  or nonAbelian but local and weakly coupled, or  a strongly-coupled, nonAbelian, nonlocal  SCFT.     
The example of the first type of the RG flow is 't Hooft's  dual Meissner effect model (assumption) of confinement  \cite{tHooftConf}.  
 Confinement  \`a la 't Hooft shown by Seiberg and Witten by an ${\cal N}=1$ perturbation of the ``monopole point" of ${\cal N}=2$  susy  $SU(2)$ gauge theories \cite{SW1,SW2}, is an explicitly realization.  

   The second type of the RG flow is the one  into nonAbelian but weakly-coupled infrared-free low-energy systems (hence,   again, ``trivial" conformal fixed points), and confinement caused by a
   relevant perturbation.  A known example is the so-called  $r$-vacua of ${\cal N}=2$ supersymmetric quantum chromodynamics (SQCD),  leading to dual Meissner effects, involving both  Abelian and nonAbelian monopoles, upon ${\cal N}=1$ adjoint-scalar mass perturbation  \cite{APS,CKM}.

Perhaps the most intriguing type of the RG flow,  from the point of view of understanding confinement in real-world QCD,  is the one which would point towards a strongly-coupled, nonAbelian, nonlocal conformal fixed points  \cite{AGK,
GiacomelliK1, GiacomelliK2, GiacomelliK3}.   Though they are the most difficult ones to analyze in general, as they  involve nonlocal, nonAbelian systems with strongly-coupled monopoles, dyons and quarks  (meaning that  the system has no Lagrangian description),     some remarkable developments  (Gaiotto-Seiberg-Tachikawa duality \cite{GST}) allow us to 
analyze the system explicitly,  and to prove confinement, 
upon appropriate  ${\cal N}=1$ perturbations.   Color confinement   in the true sense  (a)    (i.e.,  without   Abelianization) is indeed  realized in these models, as has been shown in  \cite{GiacomelliK1, GiacomelliK2, GiacomelliK3}.      
See Fig.~\ref{Confinement}.

\subsection{A final reflection}

   The RG flow of an asymptotically-free gauge theory with massless matter fields is directed towards an infrared-fixed point theory.  When a relevant perturbation is present (either in the UV theory, such as a mass term, or produced by the system dynamically, in the form of  composite scalar fields)  the RG flow may get deviated at the IR end of the trajectory (Fig.\ref{figRGflow}), leading to a confining vacuum.
  A conformal theory (CFT), a scale invariant theory,  and confinement (generation of mass scale and breaking of dilatation symmetry) might look at first sight diametrically opposite,  conceptually. 
   How can they be close to (or deformed into)  each other?
   
   There are at least two precise senses in which they can indeed be ``close to each other".  The first is that one of them may go into another when the parameters of the theory is 
   varied (such as the number of the flavors, or the mass of the matter fermions). Namely they can be close to each other in the space of theories.  
   In the standard $SU(3)$ QCD with $N_F$ massless quark flavors, confining and conformal vacua are believed to be separated by an (unknown) critical flavor number $N_F^*$. 
     
   More significantly, we learn from the analysis of the supersymmetric theories 
  that {\it   the same degrees of freedom (monopoles, dyons and quarks) describe  both  the infrared fixed-point CFT  and the nearby (perturbed by some relevant operators) vacuum in confinement phase. }
  
  In the real-world $SU(3)$ QCD,  with two nearly massless quarks,  we may exclude Abelian or nonAbelian infrared-free phases, for the lack of any phenomenological evidence.
  Among the different RG flows (Fig.~\ref{figRGflow}), then,  the only one which seems plausible is the one towards the confining vacuum lying near a nonAbelian, strongly-coupled nonlocal conformal fixed point.  It is possible that a phenomenon very similar to the one studied in \cite{AGK}-\cite{GiacomelliK3} is indeed realized in the standard  (${\cal N}=0$)  QCD,  even though our ability  of exhibiting the dynamical details is, for the moment, limited to  the  ${\cal N}=2$  supersymmetric cousins.

\begin{figure}
\begin{center}
\includegraphics[width=6in]{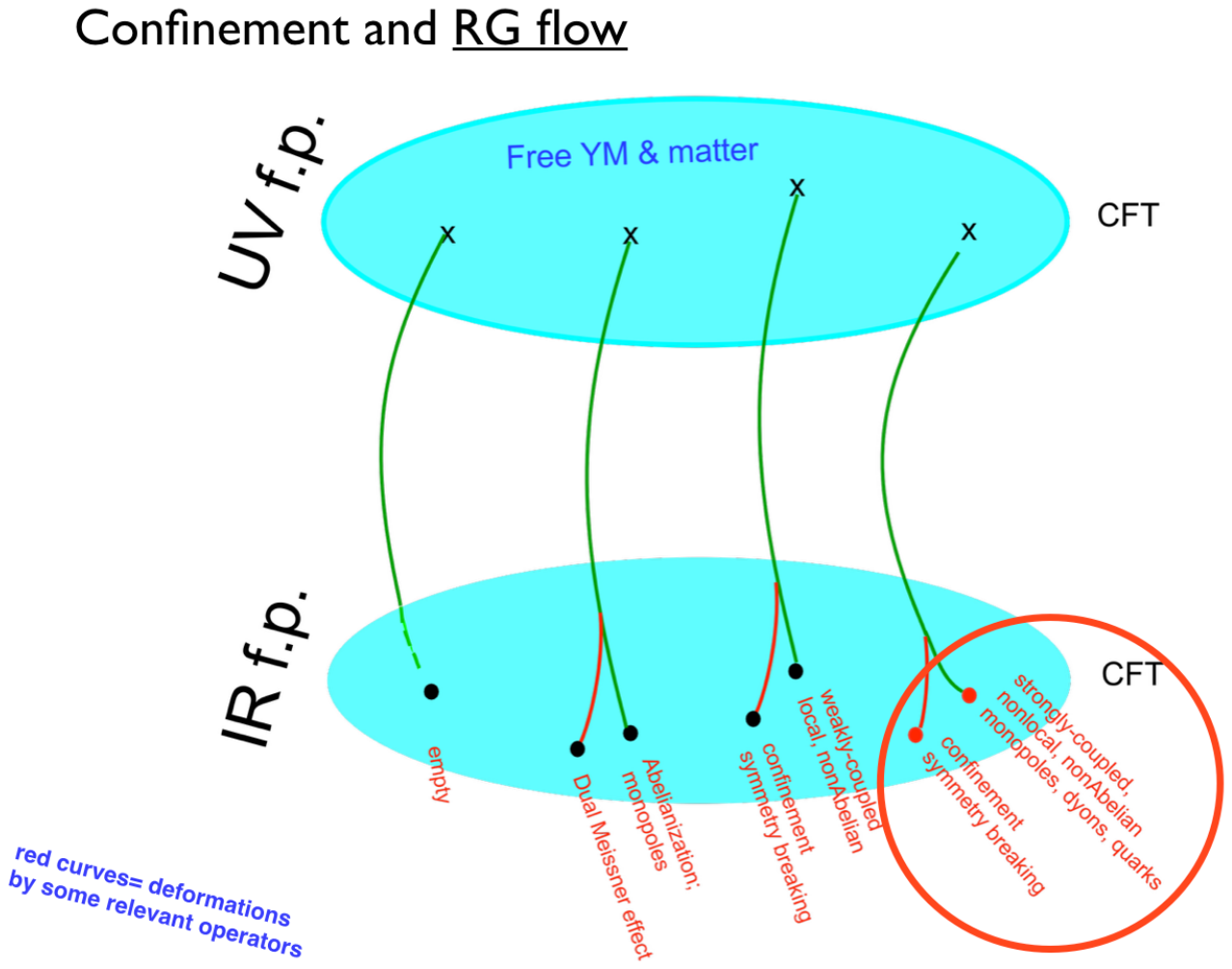}
\caption{\footnotesize  Various types of renormalization group (RG)  flows. The IR fixed points can be trivial, Abelian free theories, nonAbelian but local theories, or a strongly-coupled, nonAbelian, and nonlocal CFT.
Different types of confinement can occur, accordingly.}
\label{figRGflow}
\end{center}
\end{figure}

\begin{figure}
\begin{center}
\includegraphics[width=6.5in]{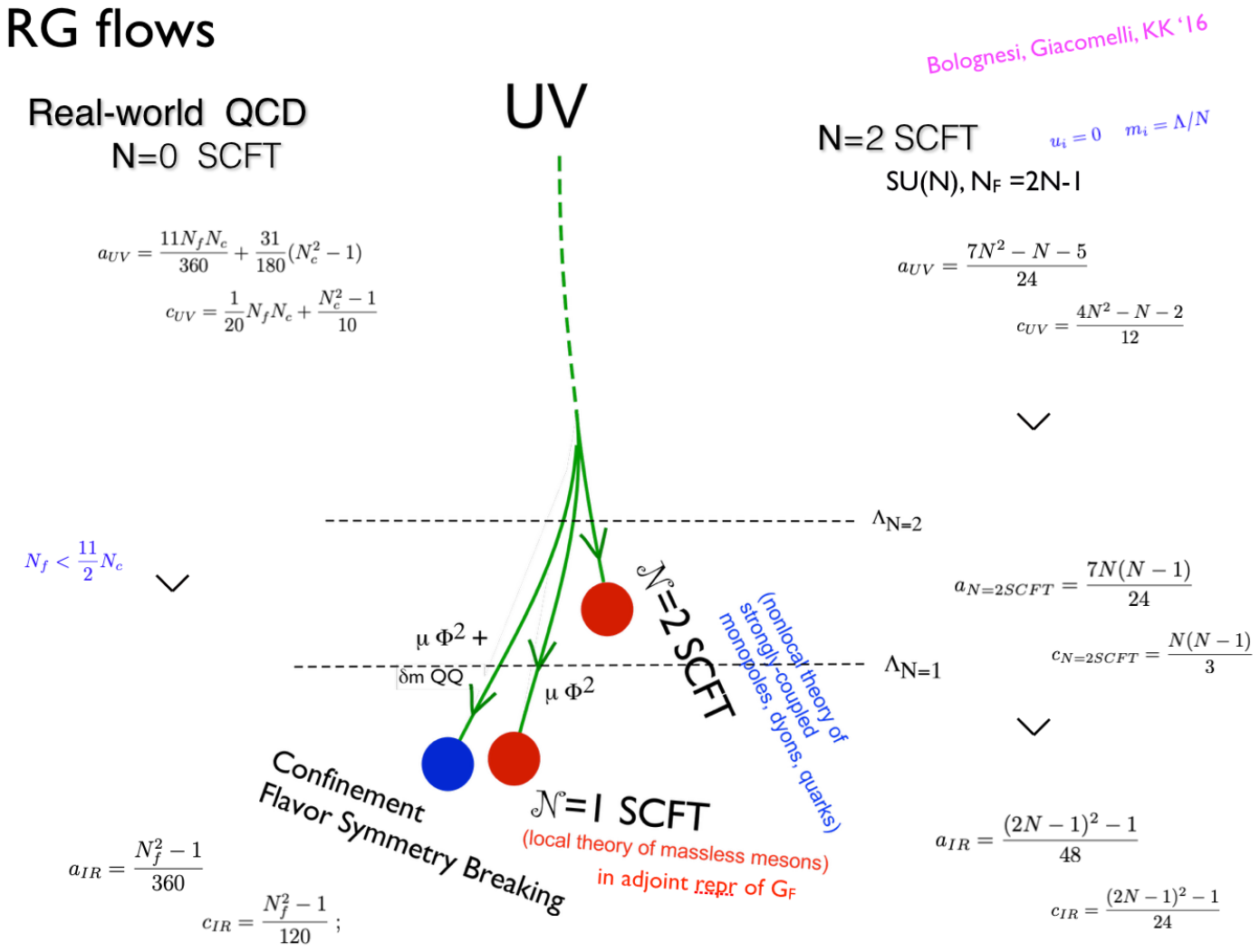}
\caption{\footnotesize   An illustration of a strongly-coupled, nonAbelian CFT  occurring in ${\cal N}=2$ SQCD  with $N_f=2N_c-1$,  deformed into a confining vacuum with relevant,  ${\cal N}=1$ perturbations.  As a side remark,  the $a$ theorem (showing the consistency of the RG flow   \cite{KomarSchw})
is illustrated in this model and in the standard, real-world QCD.   }
\label{Confinement}
\end{center}
\end{figure}

\section*{Acknowledgments}
K.K. thanks all his collaborators  in  \cite{BKS}-\cite{GiacomelliK3}.   He also acknowledges useful discussions with N. Dorey,  H. Murayama, E. Poppitz, M. Shifman,  D. Tong and G. Veneziano, at various stages of these investigations.
The work  reported here has been supported by the INFN special initiative grant, ``GAST" (Gauge and String Theories).

\end{document}